\documentclass[aps,reprint,groupedaddress]{revtex4-1}
\usepackage{graphicx,bm}
\usepackage{color}
\usepackage{blindtext}
\usepackage{amsmath,amsbsy,amssymb,amsthm,array,physics}

\begin{document}

\title{U(1) dynamics in neuronal activities}

\author{Chia-Ying Lin}
\affiliation{Department of Physics, National Tsing Hua University, Hsinchu 300044, Taiwan}

\author{Ping-Han Chen}
\affiliation{Department of Physics, National Tsing Hua University, Hsinchu 300044, Taiwan}

\author{Hsiu-Hau Lin}
\email[]{hsiuhau.lin@phys.nthu.edu.tw}
\affiliation{Department of Physics, National Tsing Hua University, Hsinchu 300044, Taiwan}

\author{Wen-Min Huang}
\email[]{wenmin@phys.nchu.edu.tw}
\affiliation{Department of Physics, National Chung Hsing University, Taichung 402204, Taiwan}
\date{April 30, 2021}

\begin{abstract}
Neurons convert the external stimuli into action potentials, or spikes, and encode the contained information into the biological nerve system. Despite the complexity of neurons and the synaptic interactions in between, the rate models are often adapted to describe neural encoding with modest success. However, it is not clear whether the firing rate, the reciprocal of the time interval between spikes, is sufficient to capture the essential feature for the neuronal dynamics. Going beyond the usual relaxation dynamics in Ginzburg-Landau theory for statistical systems, we propose the neural activities can be captured by the U(1) dynamics, integrating the action potential and the ``phase" of the neuron together. The gain function of the Hodgkin-Huxley neuron and the corresponding dynamical phase transitions can be described within the U(1) neuron framework. In addition, the phase dependence of the synaptic interactions is illustrated and the mapping to the Kinouchi-Copelli neuron is established. It suggests that the U(1) neuron is the minimal model for single-neuron activities and serves as the building block of the neuronal network for information processing.
\end{abstract}
\maketitle

\section{Introduction}

\begin{figure*}
\centering
\includegraphics[width=0.65\textwidth]{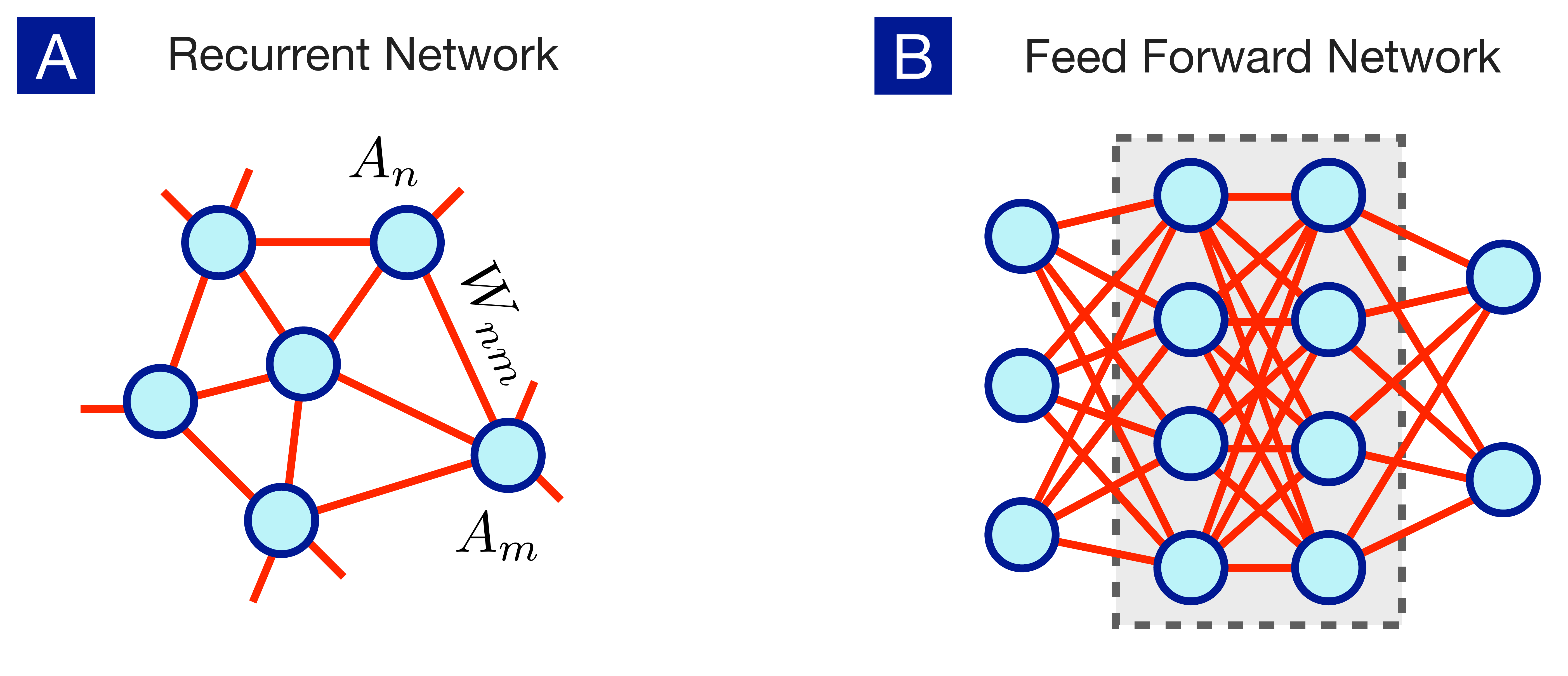}
\caption{Neurons with different network structures: (A) Recurrent neuronal network often observed in biological nerve systems. (B) Feed forward network with input, output and hidden layers, largely put to practice in deep learning.}
\end{figure*}

Human beings rely on their nervous systems to detect external stimuli and take proper reactions afterwards. Neurons are the fundamental units in the nervous system and deserve careful and thorough characterizations for their responses to external stimuli\cite{Galizia2013,Gerstner2014,Izhikevich2007,Dayan2001}. However, because neurons display considerable diversity in morphological and physiological properties, it is rather challenging to pin down the essential degrees of freedom even when studying the single-neuron dynamics. While the neural encoding and decoding\cite{Dayan2001} are not fully understood yet, the action potentials, or the spikes, of the neurons upon external stimuli are the apparent means to pass the information onto the nervous system. In consequence, the firing rate of the neuronal spiking is often used for data analysis and modeling\cite{Wilson1972,Ermentrout1994,Gerstner1995,Shriki2003,Ostojic2011}. The firing-rate models are appealing due to their simplicity and accessibility for numerical simulations.

Although the firing-rate models are helpful descriptions for neural circuits, it is not clear whether the neuronal spiking alone is sufficient to capture the essential features of neuronal activities. At the microscopic scale, the electric activities of a single neuron arise from a variety of ionic flows passing the relevant ion channels embedded on the membrane of the nerve cell \cite{Choe2002,Naylor2016}. The conductance-based approach, such as the Hodgkin-Huxley model \cite{Hodgkin1952}, provides an effective description for the emergence of action potentials by the inclusion of gating variables of the ion channels involved\cite{Gerstner2014}. Above some current threshold, the neuronal spikes starts to appear. Hodgkin proposed to classify the neurons into type I or type II depending on whether the firing rate changes continuously or discontinuously above the current threshold\cite{Hodgkin1948}. While the conductance-based model with ion-channel dynamics explains the emergence of the neuronal spikes and provides a better description for biological details, it blurs the priority of various degrees of freedom in a single neuron, rendering a clear understanding of neuronal dynamics intractable.

The dynamical transitions in a single neuron above the current threshold posts another challenge. Both the biologically based models, such as the Hodgkin-Huxley model and its generalizations, or the reduced neuron models, including the Ermentrout-Kopell model \cite{Ermentrout1986}, the FitzHugh-Nagumo model \cite{Fitzhugh1961}, the Izhikevich model\cite{Izhikevich2003,Izhikevich2004} and so on, exhibit rich types of dynamical phase transitions above the firing threshold. There are three major types of dynamical phase transitions found in single-neuron dynamics: saddle-node on invariant circle (SNIC), supercritical Hopf bifurcation and subcritical Hopf bifurcation\cite{Gerstner2014,Izhikevich2007}. Ermentrout\cite{Ermentrout1996} showed that the type I neurons undergo a SNIC dynamical transition at the threshold, while Izhikevich\cite{Izhikevich2007} pointed out that type II neurons may go through all three different bifurcations. It would be great to build a theoretical framework, capturing the essential degrees of freedom for spiking neurons and incorporating all types of dynamical phase transitions systematically.

In this Article, we propose the essential degrees of freedom for a spiking neuron are the membrane potential and the temporal sequence. It is rather remarkable that both can be integrated into a unified theoretical framework described by a single complex dynamical variable and the U(1) dynamics emerges naturally. The real part of the complex dynamical variable represents the potential of the neuron and the phase describes the temporal sequence during the firing process. When describing the neuronal dynamics of the complex dynamical variable, our U(1) neuron model not only reproduces the action potentials from the Hodgkin-Huxley neuron but also capture the gain function of the firing rate in response to the external current. 

It is known the classification of spiking neurons is closely related to the bifurcation of neuronal dynamics. The gain function of the type I neuron is continuous at the current threshold while the type II neuron exhibits a discontinuous jump at the threshold. The U(1) neuron described by the complex variable provides a natural explanation for the classification. The bifurcations on the complex plane, either in radial or phase directions, leads to various transitions among resting, excitable, firing states of a single neuron. In short, the U(1) neuron not only captures the single-neuron activity upon external stimuli, but also provides a coherent understanding for the dynamical phase transition between different types of neuronal activities. 

The major impact of the U(1) neuron is not to provide a realistic description of a spiking neuron (although it can be done as shown in the later paragraphs), just like the Fermi liquid theory is not targeting on providing a precise quantitative description for metals. The key is to grab the essential features in neuronal dynamics so that model building for different purposes can be facilitated with these ingredients. For instance, within the U(1) neuron description, we find the phase dynamics of spiking neurons is nonuniform during the firing process and the firing rate is thus dictated by the bottleneck (phase regime with smaller angular velocity). In addition, it is known that the neuron reacts differently when stimulated in different firing processes. Going beyond the usual Kumamoto-like interactions, the U(1) neuron framework provides a systematic approach to describe the phase dependence of the synaptic interactions on the presynaptic and postsynaptic neurons. With the phase dependence in mind, the refractory effect can be incorporated seamlessly into the U(1) neuron framework. In fact, we show that the Kinouchi-Copelli neuronal network is equivalent to the discrete version of the U(1) neuronal network and the spontaneous asynchronous firing (SAF) state, crucial for information processing, can be realized when the synaptic strength is strong enough.

The remainder of the paper is organized as follows. In Section II, we first compare the artificial and biological neuronal networks and point out the importance of neuronal dynamics. In Section III, we discuss the mode-locking phenomena in biological neurons. In Section IV, we go beyond the usual Ginzburg-Landau theory and construct the theoretical foundation for the U(1) neuron. In Section V, we demonstrate how the Hodgkin-Huxley neuron can be described with the theoretical framework of the U(1) neuron. In Section VI, we reveal the importance of phase dependence in the synaptic interactions and establish the equivalence of the Kinouchi-Copelli neuron and the discrete version of the U(1) neuron. Finally, we extend the single-neuron approach to neuronal network and show that the spontaneous asynchronous firing phase, beneficial for information processing, in the Kinouchi-Copelli neuronal network.

\begin{figure*}
\centering
\includegraphics[width=0.85\textwidth]{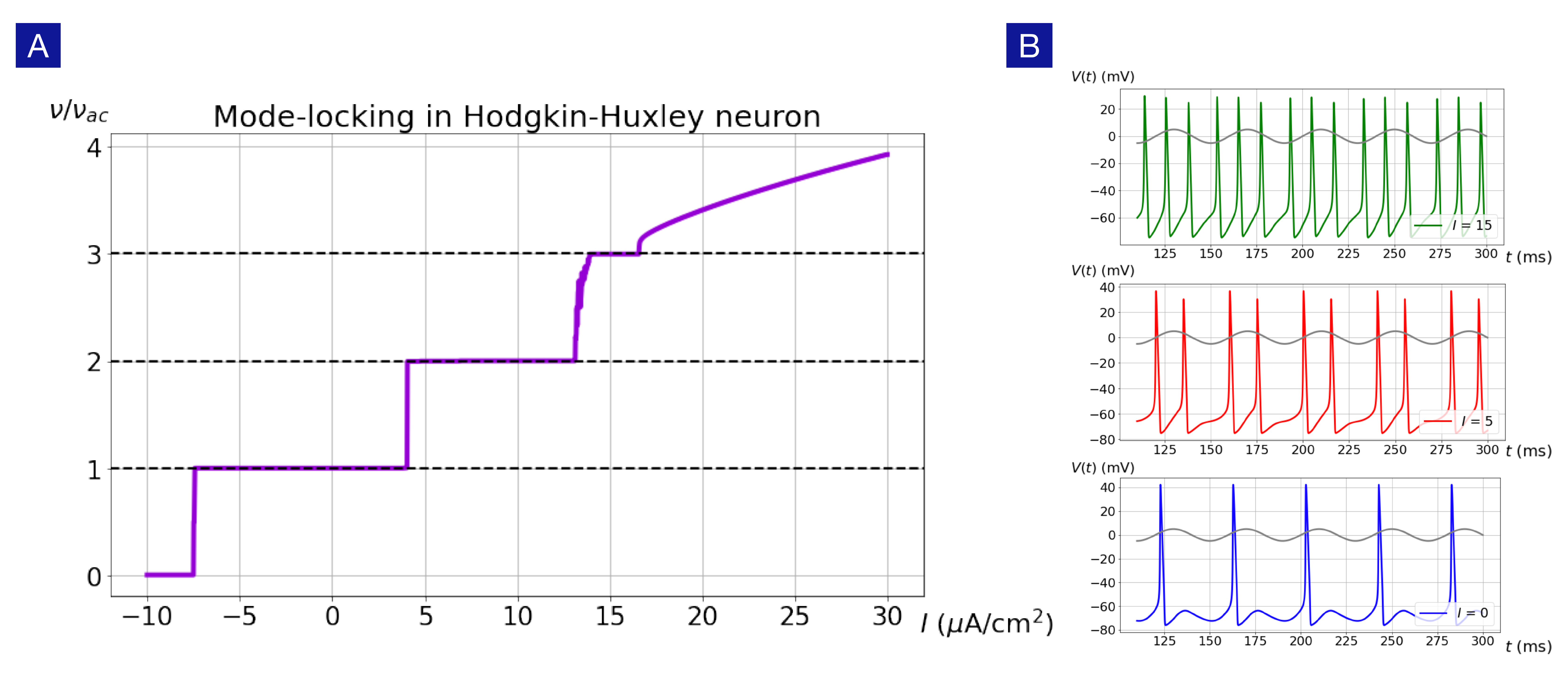}
\caption{Mode-locking in the Hodgkin-Huxley neuron. (A) The firing rate of the Hodgkin-Huxley neuron exhibits roust plateaus at $\nu = n \nu_{\rm ac}$, where $n$ is an integer, in the presence of an external ac drive with frequency $\nu_{\rm ac} = 25$ Hz. (B) Action potentials at $I = 0, 5, 15$ $\mu$A/cm$^2$ exhibits different mode-locking behaviors with $\nu/\nu_{\rm ac} = 1, 2, 3$ respectively.
}
\end{figure*}

\section{Artificial and Biological Neurons}

We first briefly explain the difference between artificial and biological neurons as shown in Figure 1. In biological nerve system, recurrent neuronal network is often found, while the artificial neural networks used in deep learning\cite{Goodfellow2016,Haykin2009} usually belong to the feed-forward type with well defined input, output and hidden layers. Despite the difference between the network structure, the neuronal dynamics within the rate-model framework is captured by the set of coupled non-linear differential equations,
\begin{eqnarray}
\tau \frac{dA_n}{dt} = -A_n + G\left( \sum_m W_{nm} A_m + I_n \right),
\end{eqnarray}
where $A_n$ denotes the firing rate of the single neuron or the activity of the neuronal population at the $n$-th node, $I_n$ is the external current injected into the $n$-th neuron and $W_{nm}$ is the synaptic weight from the $m$-th neuron to the $n$-th neuron. The relaxation time $\tau$ is the typical time scale for a single neuron returning to its resting state when the external stimulus is turned off. The gain function $G$ relates the firing rate with the stimuli, including the external current and the synaptic currents from connected neurons. The variety of individual neurons is ignored here for simplicity but can be included without technical difficulty.

Even with the simple relaxation dynamics, the coupled non-linear differential equations are already extremely complicated. Because the synaptic dynamics (how $W_{nm}$ evolves with time) is typically much slower than the neuronal dynamics, the stationary solution plays a significant role in some cases,
\begin{eqnarray}
A_n = G\left( \sum_m W_{nm} A_m + I_n \right)
\end{eqnarray}
Note that the true dynamics drops out completely in the above relations. In the feed-forward network, these are exactly the relations between input and output neurons commonly used in the artificial neural network. The synaptic weight $W_{nm}$ can be adjusted by employing appropriate algorithm to minimize the cost function but the evolution of $W_{nm}$ at different epochs does not represent the true dynamics of the neuronal network.

The deep neural network\cite{Goodfellow2016} enjoys great success in recent years and makes strong impacts in many areas in science and technology. However, as explained in the previous paragraph, it explains the slower process such as learning but does not include the reactive information processing at the shorter time scale. If the neuronal dynamics is properly included, shall the neuronal network with both types of dynamics exhibits different class of intelligence? The first step to answer this important question is to capture the essential features in neuronal dynamics before constructing the network with complicated structures. It will become evident later that the activity (firing rate) $A_n$ is insufficient to describe the dynamics of a single neuron and more degrees of freedom must be included to account for proper synaptic interactions.

\section{Mode-Locking in a Single Neuron}

Mode locking\cite{Jensen1984,Flaherty1978,Fletcher1978} is a common phenomena in physical and biological systems\cite{Guevara1981,Aihara1986,Takahashi1990,Gray1996,Szucs2001} with non-linear dynamics and may play an important role in neural information processing. In a wide variety of neuron models including Hodgkin-Huxley model, FitzHugh-Nagumo model, Izhikevich model and some integrate-and-fire models, the firing rate $\nu$ is locked to the integer multiples of the oscillatory frequency $\nu_{\rm ac}$ of the external ac current. As shown in Figure 2, the Hodgkin-Huxley neuron shows robust mode-locking behavior in the presence of the time-dependent current stimulus,
\begin{equation}
I(t) = I + I_{\rm ac}(t) = I + I_{\rm ac} \sin (2\pi \nu_{\rm ac}t),
\end{equation}
where $I$ and $I_{\rm ac}$ represent the strengths of dc and ac currents respectively. It is rather remarkable that, even with a moderate $I_{\rm ac}$, the gain function of the Hodgkin-Huxley neuron changes drastically and robust firing-rate plateaus appear with $\nu = n \nu_{\rm ac}$, where $n=0,1,2,\cdots$. The action potentials on the firing-rate plateaus indicate clear mode-locking behaviors as shown in Figure 2(b).  

The mode-locking phenomena have been known in the neuroscience community for quite a long time but its deeper implication seems neglected. First of all, the mode-locking phenomena provide a stable method to transfer information between neurons in the presence of stochastic noises. Furthermore, the mode-locking phenomena manifest the underlying {\em non-linear} phase dynamics, or the so-called U(1) dynamics.

\begin{figure}
\centering
\includegraphics[width=0.75\columnwidth]{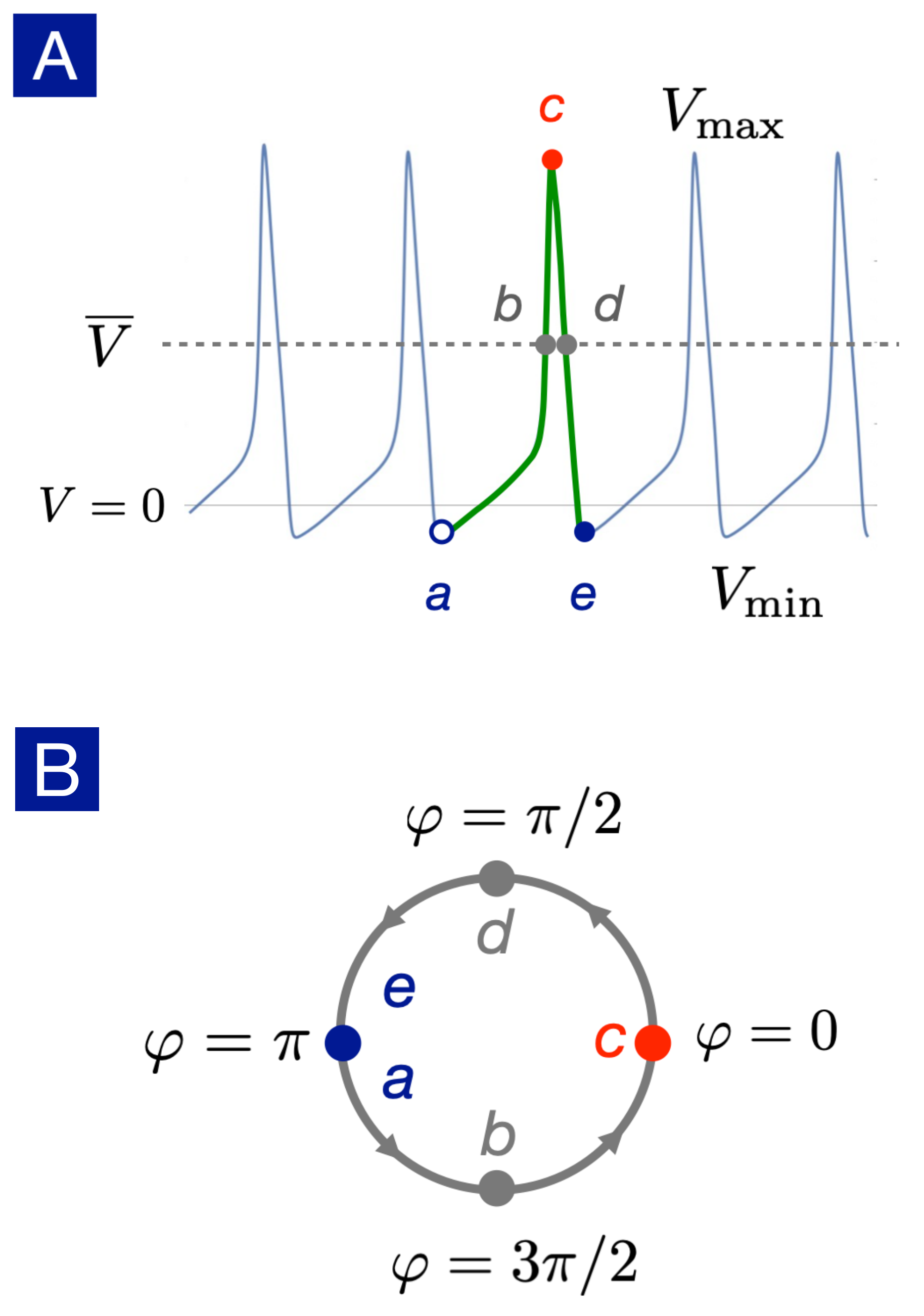}
\caption{Definition of the U(1) phase. (A) The action potential of a spiking neuron is marked by its potential minimum $V_{\rm min}$ (point $a, e$), average $\overline{V}$ (point $b, d$) and maximum $V_{\rm max}$ (point $c$). (B) The points from $a$ to $e$ corresponds to the U(1) phase $\varphi = \pi, 3\pi/2, 0, \pi/2, \pi$ respectively. A complete firing process can thus be viewed as a winding process of $\Delta \varphi = 2\pi$ in the phase dynamics.}
\end{figure}

To unveil the underlying phase dynamics, we introduce a complex dynamical variable to describe the neuronal dynamics,
\begin{equation}
z(t) = r(t) \exp[i\varphi(t)],
\end{equation}
where $r(t)$ denotes the firing amplitude and $\varphi(t)$ represents the phase during the firing process. We propose the essential degrees of freedom for a spiking neuron are the membrane potential $V(t) = r(t) \cos \varphi(t) $ and the phase $\varphi(t)$ as shown in Figure 3. The phases $\varphi=0, \pi$ correspond to the potential maximum and minimum respectively. When a spiking neuron completes a full firing cycle, it can be viewed as a winding process of $\Delta \varphi = 2\pi$ in the phase dynamics. 

In general, the angular velocity $\Omega(\varphi)$ is not constant and depends on the phase,
\begin{equation}
\frac{d\varphi}{dt} = \Omega(\varphi).
\end{equation}
As a demonstrating example, we extract the nonuniform phase dynamics of the Hodgkin-Huxley neuron with our U(1) neuron framework. The phase dynamics reveals lots of interesting features as shown in Figure 4. First of all, the angular velocity $\Omega(\varphi)$ is nonuniform, showing a bottleneck (small angular velocity) starting around $\varphi=\pi$ and a whirlwind (large angular velocity) slight below $\varphi = 0$. A non-trivial correlation between membrane potential $V(t)$ and the phase $\varphi(t)$ is thus established: the phase dynamics is fast near potential maximum while it slows down near potential minimum.  

Upon changing the external current stimulus, the overall shape of $\Omega(\varphi)$ remains more or less the same, indicating the nonuniform phase dynamics we found here is an intrinsic property of the Hodgkin-Huxley neuron. Zooming into the finer differences caused by different current stimuli, a larger current increases the angular velocity slightly near the bottleneck regime while slows down the phase dynamics around the whirlwind regime. It means that an increase of the injected current suppresses the non-uniformity of the angular velocity.

\begin{figure}
\centering
\includegraphics[width=\columnwidth]{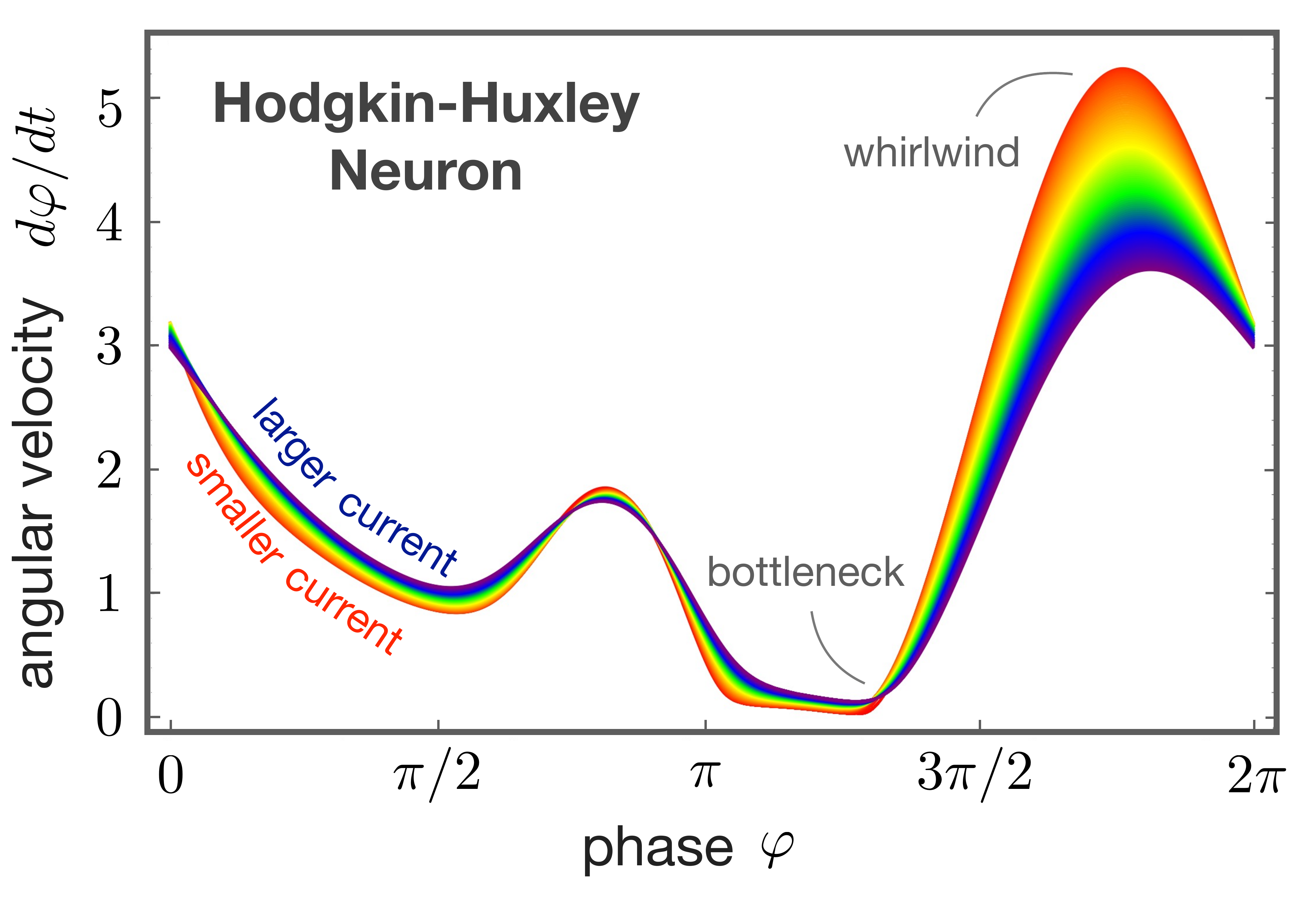}
\caption{Nonuniform phase dynamics in the Hodgkin-Huxley neuron. The angular velocity $\Omega(\varphi)$ exhibits a highly non-trivial dependence on the U(1) phase. In the regime of current stimuli from 6.3 $\mu$A/cm$^2$ (red) to 46.3 $\mu$A/cm$^2$ (blue), one finds the overall shape remains more or less the same. However, because the firing rate is dictated by the bottleneck regime, a slight increase of the angular velocity here will boost up the firing rate significantly.
}
\end{figure}

\begin{figure*}
\includegraphics[width=\textwidth]{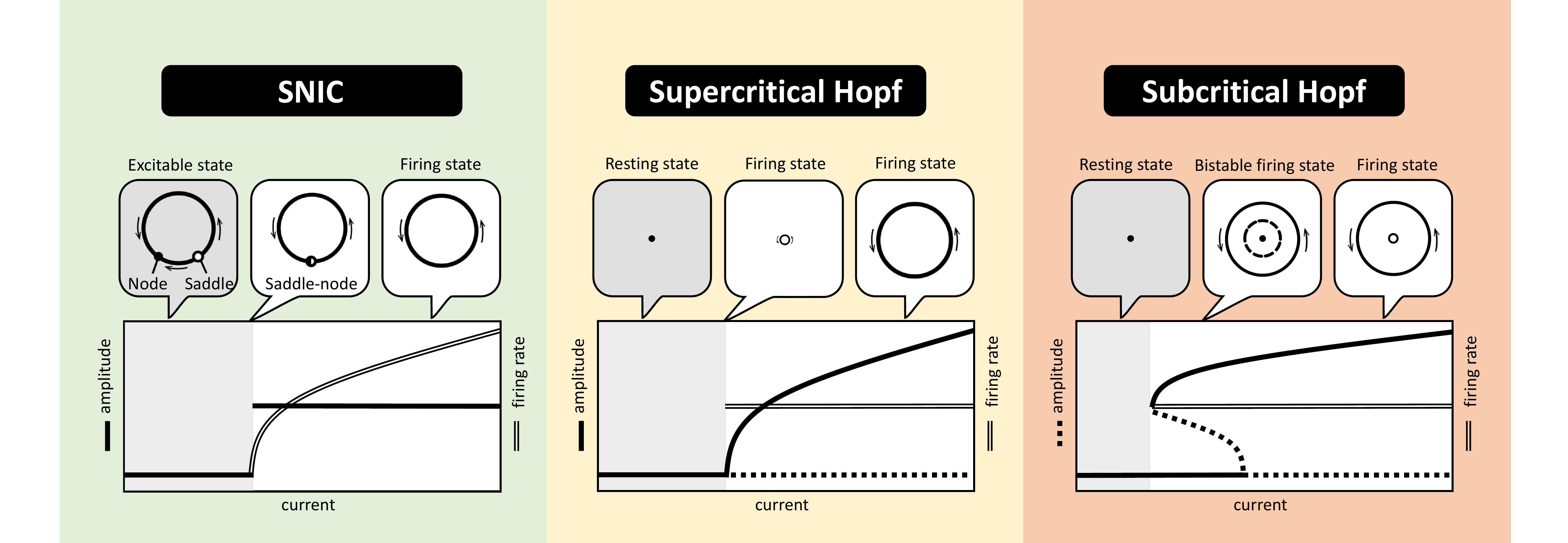}
\caption{Dynamical phase transitions in the U(1) neuron: SNIC, supercritical Hopf, subcritical Hopf. The upper panels present the topological structures of the limit cycles and the fixed points in the vicinity of the dynamical phase transitions. The firing amplitude and rate of the neuron versus the external current stimulus are illustrated in the bottom panels.}
\end{figure*}
The firing rate $\nu$ can be determined from the phase dynamics as well. Because the complete firing cycle corresponds to a $2 \pi$ phase winding, the firing rate of a spiking neuron is
\begin{equation}
\nu = \frac{1}{\Delta t_f} = \left[ \oint \frac{d\varphi}{\Omega(\varphi)} \right]^{-1},
\end{equation}
where $\Delta t_f$ is the time duration for adjacent firing events. From the above rate-phase relation, it is clear that the firing rate is dominated by the angular velocity in the bottleneck regime. Thus, a slight increase of $\Omega(\varphi)$ in the bottleneck gives rise to a significant upsurge in the firing rate $\nu$. On the other hand, the relatively large changes in the whirlwind regime for different current stimuli are irrelevant to the firing rate.

The nonuniform phase dynamics revealed by the U(1) neuron framework presents a natural explanation for the observed mode-locking phenomena. It is worth emphasizing that a uniform angular velocity $\Omega(\varphi) = \omega$ can be gauged away by redefining the complex dynamical variable $z(t) \to z(t) \exp(-i\omega t)$ but the phase-dependent angular velocity $\Omega(\varphi)$ is an intrinsic property of the neuron and cannot be gauged away. The simple yet stimulating findings presented in Figure 4 encourage us to go beyond the rate models and to integrate both membrane potential $V(t)$ and the phase $\varphi(t)$ together into a coherent theoretical framework.

\section{The U(1) Neuron}

Extending the conventional Ginzburg-Landau theory for the complex dynamical variable $z(t)$, its dynamical equation contains two parts,
\begin{equation}
\frac{dz}{dt}=-\pdv{L(z,\overline{z})}{\overline{z}}+iR(z,\overline{z})z,
\label{eq:U(1)_model}
\end{equation}
where the Lyapunov function $L(z,\overline{z})$ is a real-valued potential with U(1) symmetry and $R(z,\overline{z})$ is another real-valued function describing the nonuniform rotation of the phase. It can be shown that the Lyapunov function $L(z,\overline{z})$ decreases throughout the temporal evolution, seeking the potential minimum representing the free energy in thermal equilibrium. Thus, the radial dynamics of the firing amplitude $r(t)$ is relatively simple and, in most cases, can be understood with the usual Ginzburg-Landau theory with slight modifications. However, the presence of the nonuniform phase dynamics $R(z,\overline{z})$ go beyond the relaxation dynamics (seeking for specific potential minima) and gives rise to interesting non-equilibrium phenomena as anticipated in the excitable neuronal systems.

To make the radial and phase dynamics explicit, one can choose the biased double-well potential in the Ginzburg-Landau theory (for both the first-order and the second-order phase transtions) supplemented with the Fourier expansion for the phase dynamics,
\begin{eqnarray}
L(z,\overline{z}) &=& \frac{v_2}{2} |z|^2 + \frac{v_3}{3} |z|^3 + \frac{v_4}{4} |z|^4,
\\
R(z,\overline{z}) &=& \omega + \frac12 \sum_{n=1}^{\infty}
( c_n z^n + \overline{c}_n \overline{z}^n).
\end{eqnarray}
Here $v_2$, $v_4$ are real while $c_n$ are in general complex. Note that, unlike the Lyapunov function, the U(1) symmetry does not hold for the phase-rotation function $R(z,\overline{z})$. Separating the complex Eq.~(\ref{eq:U(1)_model}) into amplitude and phase parts, the dynamical equations read
\begin{eqnarray}
\frac{dr}{dt} &=& F(r) = v_2 r +v_3 r^2 + v_4 r^3,
\\
\frac{d\varphi}{dt} &=& \Omega(\varphi) = \omega+\sum_{n=1}^{\infty} |c_n| r^n \cos(n\varphi+\phi_{n}).
\end{eqnarray}
where $|c_n|$ and $\phi_{n}$ are the amplitudes and phases of the complex numbers $c_n$. Note that the radial dynamics contains no phase dependence and $F(r)$ serves as the conservative force driving the firing amplitude to the potential minima at $r=r^*$ satisfying the equilibrium condition $F(r^*) =0$. The phase dynamics can be rather unconventional because the angular velocity $\Omega(\varphi)$ is nonuniform\cite{Matthews1990}, a direct consequence from the spike-like action potential.

The above dynamical equations for firing amplitude and phase provides a coherent understanding of different types of dynamical phase transitions in spiking neurons. Three major types of dynamical phase transitions are illustrated in Figure 5: the saddle-node-onto-invariant-cycle (SNIC) transition is associated with bifurcations in the phase dynamics, while the supercritical and subcritical Hopf transitions are driven by bifurcations in the radial dynamics for the firing amplitude. While these dynamical phase transitions are found in various neuron models in the literature, it is rather satisfying that all three types of transitions emerge naturally within the U(1) neuron framework.

Let us focus on the SNIC transition first. As shown in Figure 5, in the presence of a finite amplitude $r^*$, the equilibrium condition $\Omega(\varphi^*)=0$ gives rise to a pair of saddle-node fixed points in phase dynamics. Below the current threshold, the saddle-node structure makes the neuron excitable. Upon increasing current injection, the saddle-node pair gets close, merges into a critical point and eventually disappears on the limit cycle. Because the limit cycle already exists below the current threshold, the firing amplitude exhibits a discontinuous jump. It is known in statistical physics that the dynamics slows down indefinitely in the vicinity of a critical point. In consequence, in near the SNIC transition, the time duration between adjacent firing events diverges, indicating a vanishing firing rate. Thus, the firing rate changes continuously across the current threshold, equating to the type I neuron.

On the other hand, in the regime where the angular velocity remains positive $\Omega>0$, the dynamical transitions are driven by the competitions among the equilibrium points in the radial dynamics determined by $F(r^*)=0$. In the supercritical Hopf transition, a stable limit cycle (non-zero $r^*$ solution) appears at the current threshold, rendering the resting state ($r^*=0$ solution) unstable. Because the limit cycle grows out from the resting state, the amplitude changes continuously across the current threshold as shown in Figure 5. Because there is no critical point involved here, the firing rate associated with the emergent limit cycle is finite in general cases. Thus, the gain function exhibits a discontinuous jump at the current threshold and corresponds to the type II neuron.

The subcritical Hopf transition arises when a pair of limit cycles appears at the current threshold. Due to the simultaneous presence of the stable limit cycle (firing state) and the stable fixed point (resting state), the neuron is bistable. As shown in Figure 5, both the firing rate and amplitude are discontinuous at the current threshold. Although neurons undergo the subcritical Hopf transition can be classified as the type II, their dynamics are more complicated in comparison with the supercritical Hopf transition where neurons are either in the resting state or the firing state.

\begin{figure}
\centering
\includegraphics[width=\columnwidth]{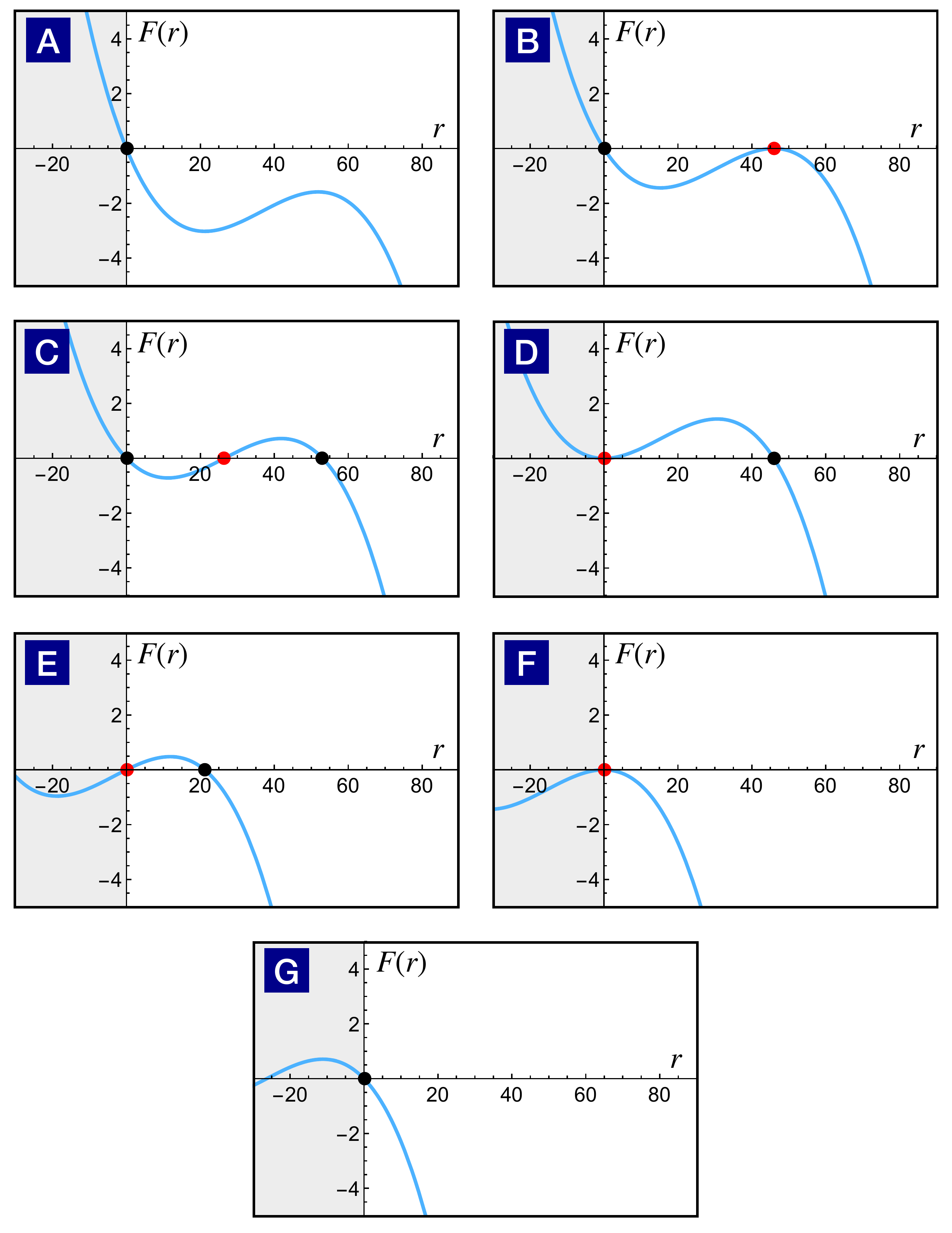}
\caption{Dynamical phase transitions driven by the $F(r)$ evolution. As the fitting function $s(I)$ changes with the external current, the cubic function slides through different configurations from (A) to (G). The black and red points indicate the radius of the stable and unstable limit cycle respectively. Configurations from (A) to (D) show the birth of a pair of limit cycles and the unstable limit cycle swallows the stable fixed point at the origin subsequently. Compared with the previous classification scheme, it belongs to the subcritical Hopf transition. Configurations from (E) to (G) illustrate the process of a stable limit cycle shrinks to an unstable fixed point, leading to a stable fixed point subsequently. This dynamical transition belongs to the supercritical Hopf.}
\end{figure}

\section{Fitting Hodgkin-Huxley Neuron}

In this section, we demonstrate how the Hodgkin-Huxley neuron can be described by the U(1) neuron in a wide range of parameter regime. The bifurcation diagram\cite{Xie2008} of the Hodgkin-Huxley neuron undergoes a subcritical Hopf transition first at the current threshold at $I=9.780$ $\mu$A/cm$^2$, followed by another supercritical Hopf transition at $I=154.527$ $\mu$A/cm$^2$. These dynamical phase transitions can be captured within the U(1) neuron framework.

Let us focus on the radial dynamics of the firing amplitude first. Because the resting state ($r^* = 0$) is always present, the potential flow in the Ginzburg-Landau theory can be constructed in the following way. Introducing a cubic function $f(r)$ with three zeroes at $r = 0, p_1, p_2$, 
\begin{equation}
f(r)= A(r-0)(r-p_1)(r-p_2),
\end{equation}
where $A$ is some positive constant. The conservative force $F(r)$ in the radial dynamics can be expressed as

The parameters $p_1$ and $p_2$ can be chosen as the amplitude of the stable and unstable limit cycle, and $a$ should be negative because of the limit of the ionic sources. Then the force in the amplitude equation is the coordinate transformation concerning the external input current $I$ that switch neurons between firing states and resting states:
\begin{equation}
\begin{split}
F(r) =&\: f[s(I)]-f[r+s(I)]
\\
=& A \Big[ p_1 p_2  - 2 p_1 s(I) - 2 p_2 s(I) + 3 s^2(I)
\\
& + 3 s(I) r - (p_1 + p_2) r + r^2 \Big]r.
\end{split}
\end{equation}
Here $s(I)$ depends on the external current and serves as a fitting function to reproduce the correct dynamical phase transitions in the targeted parameter regime.

The solutions for $F(r^*)=0$ describes the fixed point or the amplitudes of the limit cycles,
\begin{equation}
\begin{aligned}
r^*_{0} =&\: 0,
\\
r^*_{\rm stable} =&  \frac{1}{2} \Bigg[p_1 + p_2 - 3 s(I)
\\
&\hspace{-20mm} + \sqrt{p_1^2 - 2 p_1 p_2 + p_2^2 + 2 p_1 s(I) + 2 p_2 s(I) - 3 s(I)} \Bigg],
\\
r^*_{\rm unstable} =&  \frac{1}{2} \Bigg[p_1 + p_2 - 3 s(I)
\\
&\hspace{-20mm} - \sqrt{p_1^2 - 2 p_1 p_2 + p_2^2 + 2 p_1 s(I) + 2 p_2 s(I) - 3 s(I)} \Bigg]
\end{aligned}
\end{equation}
Note that only non-negative solutions are physical because the firing amplitude cannot be negative. The $r^*=0$ solution is always present as anticipated. The other two solutions appear in pair and represent a pair of stable and unstable limit cycles. 

\begin{figure}
\centering
\includegraphics[width=\columnwidth]{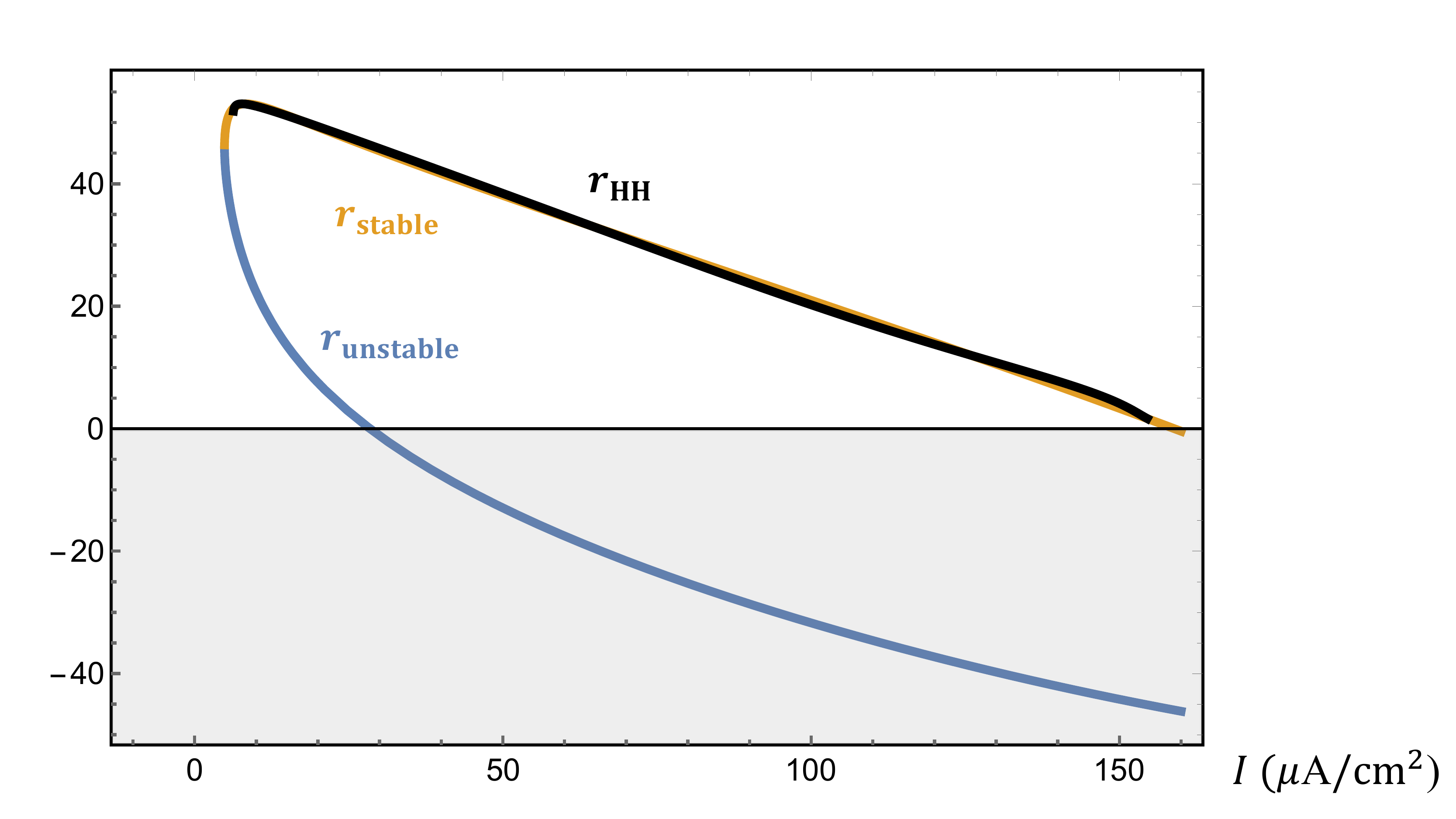}
\caption{Fitting the firing amplitude of the Hodgkin-Huxley neuron. The firing amplitude of the effective U(1) neuron (orange) fits that of the Hodgkin-Huxley neuron (black) rather well. The firing amplitude of the unstable limit cycle (light blue) is also shown for reference.}
\end{figure}
\begin{figure}
\centering
\includegraphics[width=0.85\columnwidth]{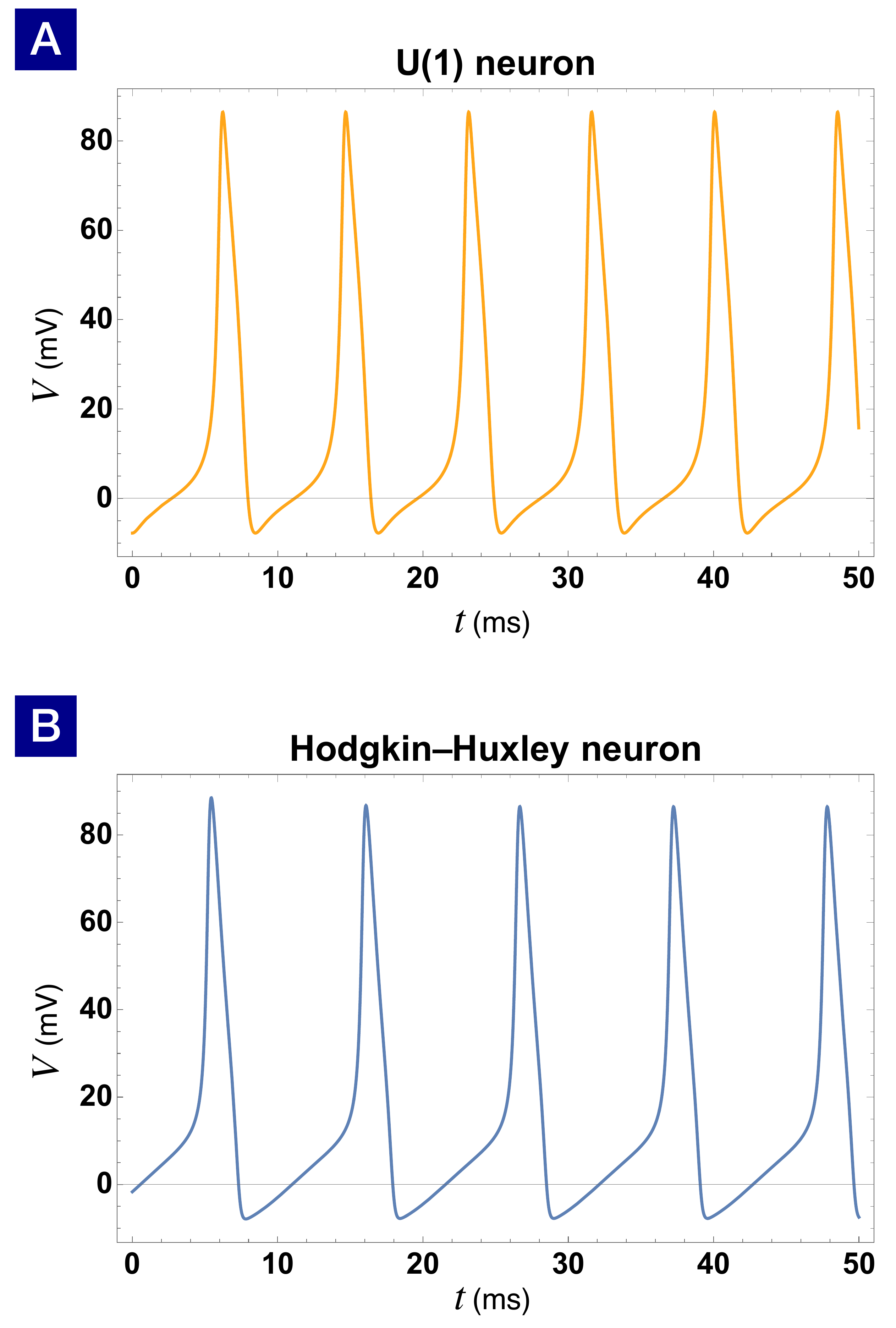}
\caption{The action potentials of (A) the effective U(1) neuron and (B) the Hodgkin-Huxley neuron are almost identical. The external current is $I=26.28$ $\mu$A/cm$^2$ for both neurons. The voltage of the resting state $V_{\rm rest} = -65$ mV is shifted to zero for visual clarity.}
\label{fig:action_potential_comparision}
\end{figure}

\begin{figure}
\centering
\includegraphics[width=\columnwidth]{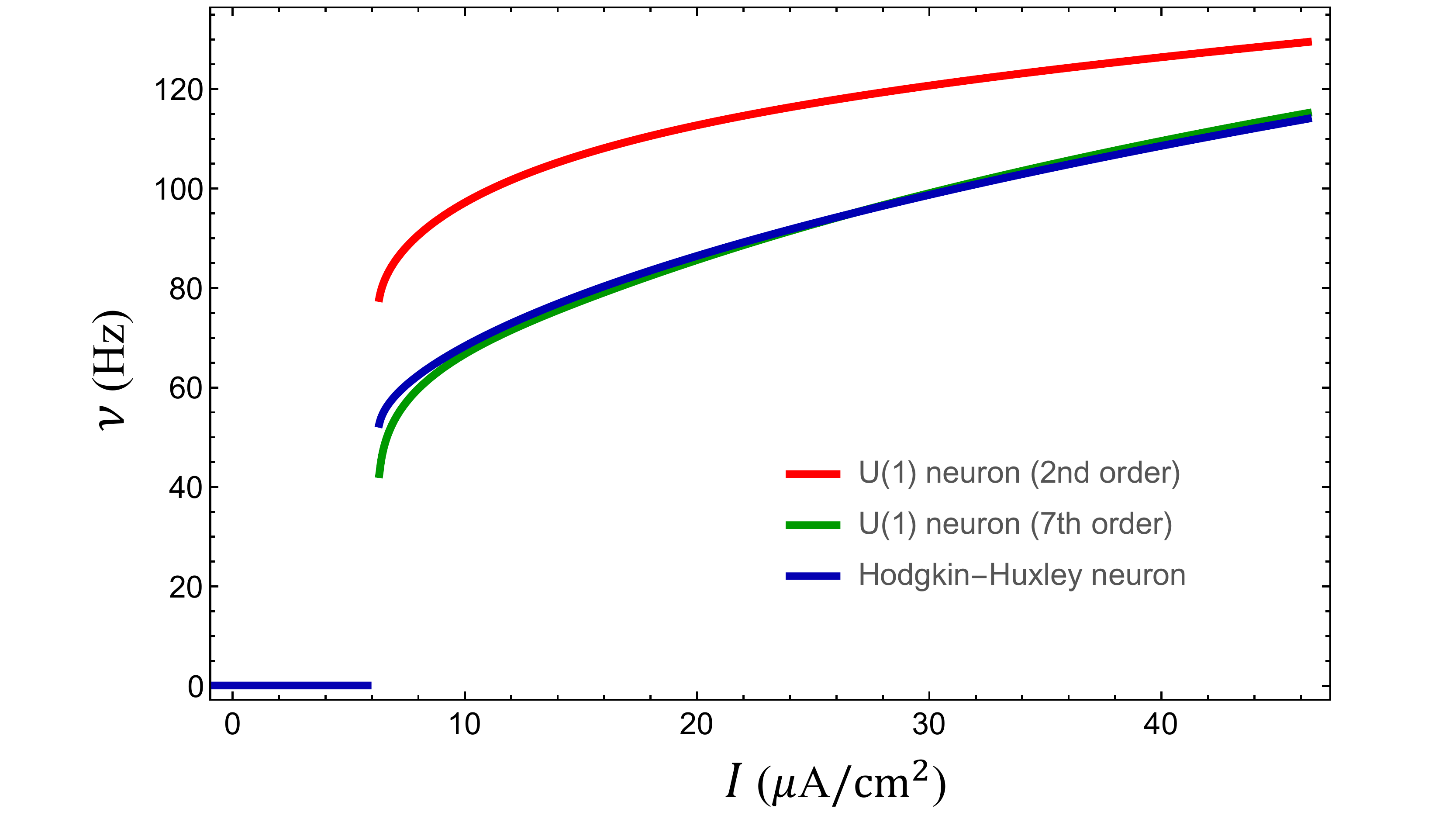}
\caption{Gain function of the U(1) neuron. The gain function of the Hodgkin-Huxley neuron (blue) is well captured by that of the U(1) neuron with Fourier expansion to the seventh order in phase dynamics. However, the U(1) neuron to the second order delivers an approximate gain function with the same trend, indicating lower-order terms already secure the qualitative behavior of the neuronal dynamics.}
\end{figure}

\begin{figure*}
\centering
\includegraphics[width=0.9\textwidth]{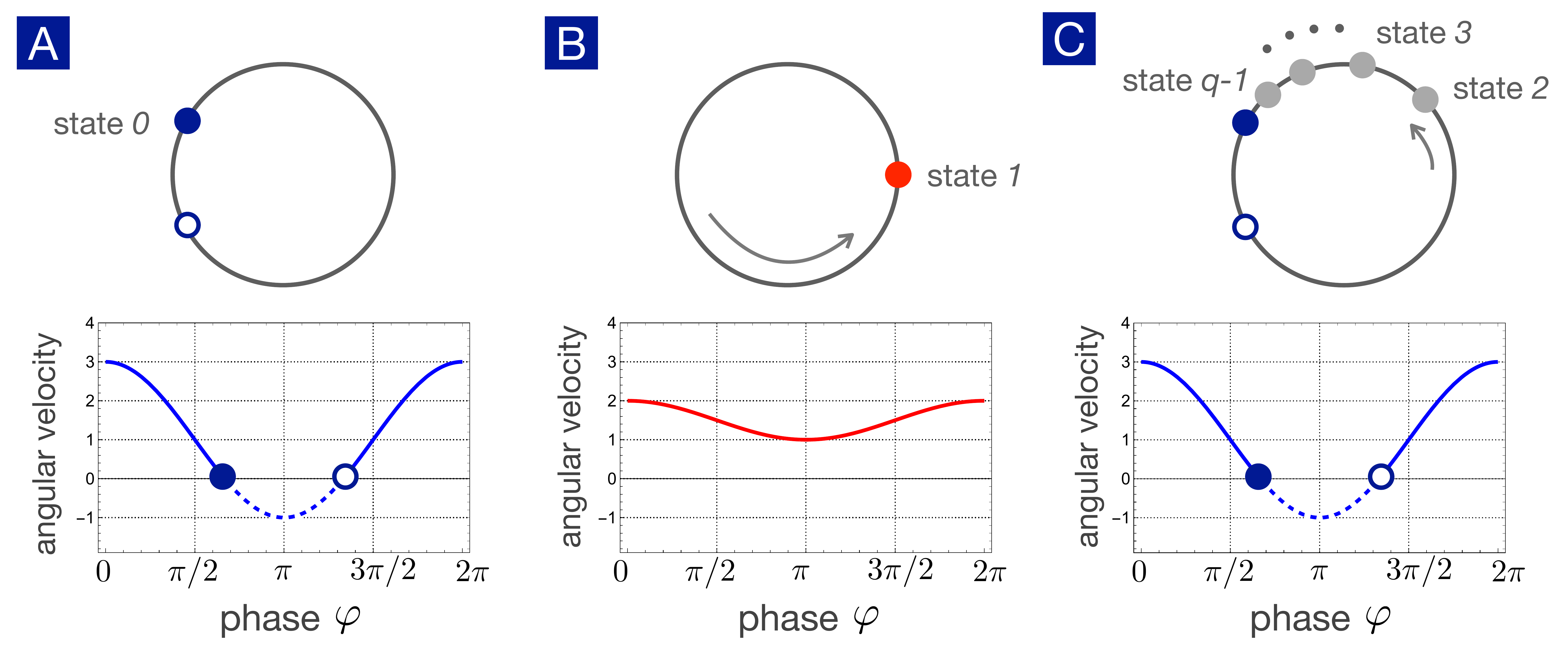}
\caption{Kinouchi-Copelli neuron as the U(1) neuron in discrete time steps. (A) Without external stimuli, the neuron is excitable with a stable fixed point (state 0) denoting the resting state. (B) In the presence of external stimuli, by either external or synaptic currents, the neuron fires and the phase swings to $\varphi=0$ with maximum potential. (C) After firing, the neuron enters the refractory states denoted by states $2,3,\cdots, q-1$ and eventually returns to the rest state.}
\end{figure*}

The remaining task to map the Hodgkin-Huxley neuron into the U(1) neuron is to equate firing amplitudes in both descriptions, $r_{\rm stable}[s(I)]= r_{\rm HH}(I)$ by numerical fitting. Keeping the lower-order terms in the fitting function $s(I)$,
\begin{equation}
s(I)=-15.4037 + 0.146443 I+ 6.70533 \ln{I},
\end{equation}
it is sufficient to match the firing amplitude rather well as shown in Figure 7. The coefficients $c_n$ in the phase dynamics can be found in numerical fitting as well. Keeping the Fourier expansion to the seventh order in phase dynamics, the action potential of the Hodgkin-Huxley neuron is almost identical to the effective U(1) neuron presented in Figure 8. 

To capture the spike-like profile of the action potential, it is necessary to include higher order terms in the phase dynamics. However, in the cases where the precise shape of the action potential is irrelevant, the precision of the Fourier expansion can be relaxed. As shown in Figure 9, the gain function of the U(1) neuron up to the seventh order in phase dynamics matches that of the Hodgkin-Huxley neuron rather well. However, the gain function up to the second order in phase dynamics still keeps the qualitative trend with reasonable compromise in quantitative precision.

\begin{figure*}
\centering
\includegraphics[width=0.8\textwidth]{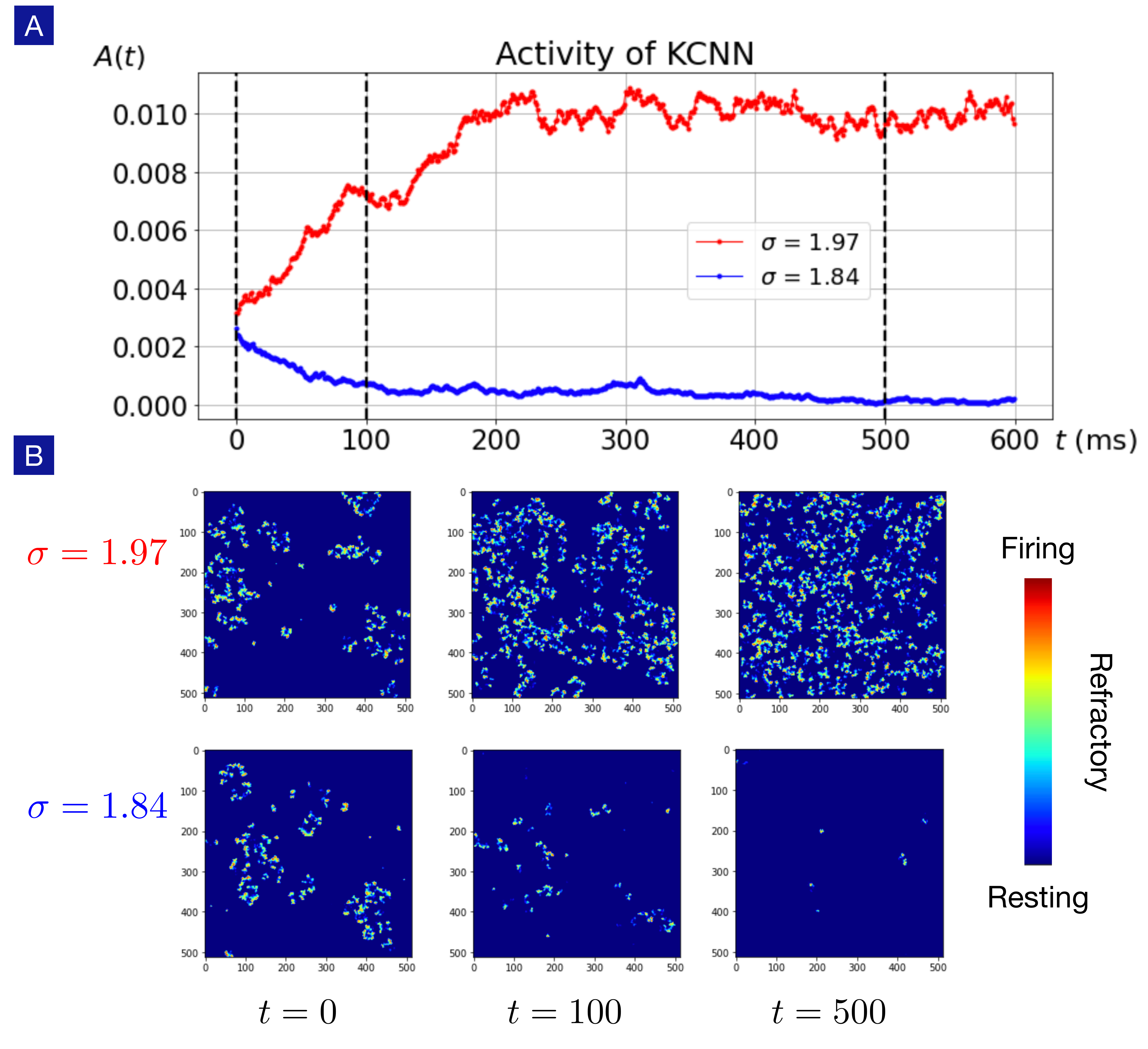}
\caption{Spontaneous asynchronous firing in Kinouchi-Copelli neuronal netwrok. (A) Given a weak neuronal activity as the initial configuration, a larger branching ratio ($\sigma=1.97$) drives the Kinouchi-Copelli neuronal network to the spontaneous asynchronous firing (SAF) phase with stable neuronal activities, while a smaller branching ratio ($\sigma=1.84$) leads to the quiescent state with diminishing neuronal activities. (B) Snapshots of the Kinouchi-Copelli neuronal network in SAF and quiescent phases.}
\end{figure*}

\section{Phase-dependent synaptic interactions}

With the established theoretical framework for a single neuron, we move on to investigate the synaptic interactions between connected neurons. As the nonuniform angular velocity in phase dynamics plays a significant role for the single neuron, we anticipate that the synaptic interactions also carry nontrivial phase dependence.

Suppose the phases of presynaptic and postsynaptic neurons are labels as $\varphi_1(t)$ and $\varphi_2(t)$ respectively. The Kuramoto interaction\cite{Acebron2005} carries the phase dependence $\cos (\varphi_1-\varphi_2)$ and tends to synchronize both neurons. In terms of the complex dynamical variables, the synaptic interaction $V_K(z_1,\overline{z}_1,z_2,\overline{z}_2)$ can be written as
\begin{equation}
V_K = K_{12} (\overline{z}_1 z_2 + \overline{z}_2 z_1),
\end{equation}
where $K_{12}$ denotes the synaptic strength of the Kuramoto interaction between the neurons. Note that the above interaction is U(1) symmetric, i.e. it is invariant under a constant phase shift for all neurons. 

The above synaptic interaction is simple and widely used in the generalized Kuramoto models\cite{Acebron2005} for studying synchronization phenomena. But, the U(1) symmetry imposes a rather unrealistic constraint on the neuronal dynamics. A constant phase shift means a temporal shift in the firing process, which is certainly not invariant for any realistic neurons. Furthermore, even though synchronization phenomena is widely observed in many biological systems, it is faulty for information processing. In fact, Parkinson's disease is correlated with excessively strong oscillatory synchronization in the brain areas such as thalamus and basal ganglia\cite{Pare1990,Nini1995,Tass2010}, while a healthy brain is asynchronous in these areas.

Let us try to model the phase dependence of the synaptic interactions from realistic neuronal properties. When the presynaptic neuron fires, it provides a synaptic current and enhances the chance for the postsynaptic neuron to fire. This process can be approximated by the phase factor $1+\cos \varphi_1$ qualitatively because it reaches the maximum at $\varphi_1=0$ (spike) and vanishes at $\varphi_1=\pi$ (hyperpolarized). It is also known that the postsynaptic neuron is sensitive to external stimuli in the hyperpolarized period while almost insensitive around the spike. Thus, there is another phase factor $1-\cos \varphi_2$ arisen from the postsynaptic neuron. Combing the phase factors for presynaptic and postsynaptic neurons together, we anticipate the synaptic interaction carries the overall phase dependence $(1+\cos \varphi_1)(1-\cos \varphi_2)$. In terms of the complex dynamical variables, the synaptic interaction $V(z_1,\overline{z}_1,z_2,\overline{z}_2)$ can be written as
\begin{equation}
V = W_{12} (2|z_1|+ z_1+\overline{z}_1)(2|z_2|-z_2-\overline{z}_2),
\label{eq:V}
\end{equation}
where $W_{12}$ denotes the strength of the synaptic interaction between the neurons. After taking the realistic neuronal properties into account, the synaptic interaction is no longer U(1) symmetric and the non-physical constraints are lifted. It is rather interesting that the above synaptic interaction is similar to the Bardeen-Cooper-Schrieffer interaction, as the U(1) symmetry is also broken in superconductors. The dynamical behaviors of neuronal networks with this type of phase-dependent synaptic interactions avoid the ultimate fate of synchronization, effective for information processing, and remain open for further investigations.

The refractory effects in spiking neurons also lead to phase-dependent synaptic interactions. Kinouchi and Copelli proposed a $q$-state neuron model to account for the observed refractory effects\cite{Kinouchi2006}. As shown in Figure 10, the resting state is labeled as state 0 and the firing state is labeled as state 1. The other states $2,3,\cdots,q-1$ are refractory and thus insensitive to external stimuli. Because the U(1) neuron provides the general theoretical framework for nonuniform phase dynamics, it is not surprising that the Kinouchi-Copelli neuron can be mapped to the U(1) neuron as well.

In the absence of external stimuli, the Kinouchi-Copelli neuron is excitable, described by a pair of saddle and node on the limit cycle as shown in Figure 10. The resting state corresponds to the stable node, labeled as state 0. When the neuron is activated by external stimuli, the angular velocity $F(\varphi)$ lifts up and the neuron fires. The phase swings to $\varphi=0$ (spike) with maximum potential, labeled as state 1. The change of the angular velocity upon external stimuli resembles the Hodgkin-Huxley neuron studied in the previous section. After firing, the neuron enters the refractory period and gradually relaxes back to the resting state. Cutting the refractory period into $q-1$ time steps of equal intervals, the refractory states $2,3,\cdots,q-1$ are defined accordingly. In short, the Kinouchi-Copelli neuron can be viewed as a discrete version of the U(1) neuron going through the SNIC transition upon external stimuli. 

Now we turn to the phase dependence of the synaptic interaction caused by the refractory effect. For simplicity, we place these neurons on a two dimensional grid to form the Kinouchi-Copelli neuronal network (KCNN). The phase dependence of the presynaptic neuron take the usual form of pulse coupling: when the presynaptic neuron fires (state 1), it gives rise to a finite probability $p$ to activate the postsynaptic neuron. In a two dimensional grid, a postsynaptic neurons are connected to 4 presynaptic neurons. Thus, it is convenient to introduce the brach ratio $\sigma$ as the sum of all activation probabilities, $\sigma = 4 p$, to parameterize the dynamical behaviors of the KCNN. The postsynaptic neuron also brings about another phase dependence due to refractory effects: only when the postsynaptic neuron is in the resting state (state 0), it can be activated to fire upon external stimuli. 

Although the phase dependence of the synaptic interactions in the KCNN is not the same as that in Eq.~(\ref{eq:V}), the broken U(1) symmetry is manifest. Therefore, spontaneous asynchronous firing (SAF) phase is anticipated when the synaptic weight is strong enough. This is indeed true. We perform numerical simulations for the KCNN with different branching ratios as shown in Figure 11. The initial configurations are randomly set with sparsely distributed activated neurons. When the branching ratio is larger than some critical value, the firing activity is enhanced and the neuronal activity is described by the SAF phase. On the other hand, when the branching ratio is smaller than the critical, the firing activities are suppressed, entering the quiescent phase with diminishing neuronal activities. The dynamical phase transition and associated statistical analysis for the KCNN will be presented in detail elsewhere.

Unlike the synchronous phase, the SAF phase can encode and decode information effectively and is of vital importance for studying neuronal networks. To achieve this goal, the phase dependence of the synaptic interactions must be properly taken into account and the U(1) neuron framework provides a natural and convenient approach to tackle this daunting challenge.

In conclusion, we propose that the rate models of spiking neurons miss the essential ingredient of neuronal dynamics and construct the U(1) neuron framework to integrate the dynamics of firing amplitude and phase coherently. We show that the dynamics of the Hodgkin-Huxley neuron, including action potentials, the gain function and associated dynamical phase transitions, is well captured within the U(1) neuron framework. Meanwhile, the nonuniform phase dynamics is unveiled with the bottleneck-whirlwind feature. We emphasize the importance of the phase-dependent synaptic interactions and map the Kinouchi-Copelli neuron as a discrete version of the U(1) neuron with refractory effects. Placing these neurons on a two dimensional grid to form a neuronal network, our numerical simulations show that the SAF phase, crucial for information processing, can be realized. Our findings suggest that the U(1) neuron is the minimal model for single-neuron activities and serves as the building block of neuronal networks for information processing.

\section*{Acknowledgements}
The authors would like to express their gratitude to Ching-Lung Hsu, Chin-Yuan Lee, Georg Northoff and Wei-Ping Pan for helpful discussions and comments. We acknowledge supports from the Ministry of Science and Technology through grants MOST 109-2112-M-007-026-MY3, MOST 109-2124-M-007-005 and MOST 110-2112-M-005-011. Financial supports and friendly environment provided by the National Center for Theoretical Sciences in Taiwan are also greatly appreciated.

\bibliographystyle{apsrev4-1}

\end{document}